# INVESTIGATION OF THE PERFORMANCE OF PHOTOVOLTAICS INSTALLED CLOSE TO RIVER

Armstrong O. Njok[1] Julie C. Ogbulezie[2]

[1]Department of Physics, Faculty of Sciences, Cross River University of Technology (CRUTECH) Calabar, Calabar, 540252, Nigeria.
[2]Department of Physics, Faculty of Physical Sciences, University of Calabar, Calabar, 540242, Nigeria.

Correspondence should be addressed to Armstrong O. Njok; Njokarmstrong@crutech.edu.ng

## ABSTRACT

The effect of temperature and relative humidity on photovoltaics close to river in calabar was studied. A KT-908 precision digital hygrometer and thermometer, and a M890C$^+$ digital multimeter were used in the process. Results obtained shows that the effect of relative humidity on current, power and efficiency are the same. The results also show that voltage remains fairly stable between 65% to 75% relative humidity. While efficiency increases with temperature up to the maximum operating temperature of the photovoltaic module before starting to decrease. Peak in efficiency was observed at a relative humidity value of 65% and module temperature of about 47$^0$C at noon.

## INTRODUCTION

Nowadays we get approximately 80% of our energy from non-renewable energy sources, e.g. fossil fuels. Pollutants and greenhouse gases increase when fossil fuels are converted into electricity or heat. Therefore the atmosphere is damaged and global warming developed. Fortunately, as the resources are limited, our dependence on fossil is close to its end [4].

Since the existence of man on earth, man has greatly relied on the use of fossil fuel for electricity and power generation. On a world scale, environmental problems including global warming due to extensive use of fossil fuels for electricity production and combustion engines have become increasingly serious in recent years. One of the viable solutions to tackling these environmental problems is by turning our attention to the use of renewable sources of energy to replace the use of fossil fuel for electricity and power generation. Solar energy which is clean, noiseless and pollution free is considered one of the most important sources among the renewable sources of energy.

Intensive efforts are being made to reduce the cost of photovoltaic cell production and improve efficiency and narrow the gap between photovoltaic and conventional power generation methods such as steam and gas turbine power generators. In order to decrease the cost of PV array production, improve the efficiency of the system and collecting more energy for unit surface area different efforts have been made [1].

Photovoltaic cells are solar energy applications and are used to convert the solar energy directly into electricity by pairs of semiconductor interacting with the effect of light [2]. The limited efficiency of the photovoltaic is the hindering reason for the widespread use of solar cells. The primary cause of the photovoltaic cell low efficiency is that it uses a small part of the energy in the solar spectrum [3].

Photovoltaic systems have been installed to provide electricity to the billions of people that do not have access to mains electricity. Power supply to remoter houses or villages, irrigation and water supply are important application of photovoltaics for many years to come. In the last decade, PV solar energy system has shown its huge potential. The amount of installed PV power has rapidly increased. Nowadays, nearly 70 GW of PV power are installed worldwide [4].

[5] Carried out a research in Sohar city on the effect of humidity on photovoltaic performance based on experimental study. Results indicated that Increasing relative humidity reduced current highly. Increasing relative humidity from 67% to 95% reduced the current by 44.44%. In spite of high relative humidity of Sohar city the PV panel produced 62% of the maximum current in the worst condition. The voltage of PV reduced with increasing relative humidity. Relative humidity has an adverse impact on solar radiation so that the resultant negative influence reflects on the PV cell output voltage. The PV panel voltage was reduced by 23.18, 24.88 and 25% for July, August, and September respectively.

[6] Studied the effect of humidity ranges between (40 to 78%). The study results indicated that there is an estimated loss of about 15-30% of the PV power. Humidity brought down the utilized solar energy to about 55-60% from just 70% of utilized energy. The reason for this reduction resulted from the basal layer of water vapor lied at the front of the solar cell directly facing the sun.

[8] Carried out an investigation in Calabar (Nigeria) about the effect of relative humidity on the performance of solar panels. Their results demonstrate that low relative humidity between 69% and 75% favors an increase in output current from solar panels, with voltage stabilizing between relative humidity values of 70% and 75%, as well increases with a decrease in relative humidity.

[9] Using a B-K Precision modules 615 Digital light instrument and PV modules in Port Harcourt carried out a research on the effect of solar flux and relative humidity on the efficient conversion of solar energy to electricity. Results obtained shows that current increases when relative humidity drops, which means low water vapor in the atmosphere, resulting to high flux which enhances high current production. Also a decrease in relative humidity leads to a decrease in voltage as well as efficiency. Their results also show that solar panel operates optimally at lower temperatures. Lower panel temperature leads to a decrease in output current and an increase in output voltage which ultimately increases the output power as well as efficiency.

[7]. Investigated the effect of temperature on the performance of a photovoltaic solar system in eastern Nigeria. The results show that there is an indirect proportionality between the power output produced by the system and the ambient temperature of the locality. Thus the application of photovoltaic technology in the conversion of solar energy to electricity is not favorable during the period of very high ambient temperature than the period of low ambient temperature. The results indicated that PV solar panels must be installed at a place where they receive more air current so that the temperature remains low while the power output remains high.

[1]. Carried out an experimental study of combining a photovoltaic system with a heating system. From their results it was concluded that the photovoltaic panel efficiency is sensitive to the panel temperature and decreases when the temperature of the panel increases. They also recommended that one way of improving the system operation is by covering the panel surface with a thin film of running water which decreases both reflection loss and temperature of the panel. Results of their work also showed that while the temperature of the panel could be controlled at a desired temperature level, the water collected at the lower end of the panel can be used as a utility for heating purposes.

## STUDY AREA

Calabar, the capital of Cross River State is located in the southern part of Nigeria, located on Latitude $4^0 57'06''$N and longitude $8^0 19'19''$E at an elevation of 42m above sea level. But the location selected for this study is on Latitude $4^0 57'38.6161''$ N and Longitude $8^0 18'58.482''$ E, it is about 500metre away from the Calabar River as shown in Fig. 1 below. This location was selected because we seek to investigate the performance of photovoltaic installed for household use in atmospheres close to the sea.

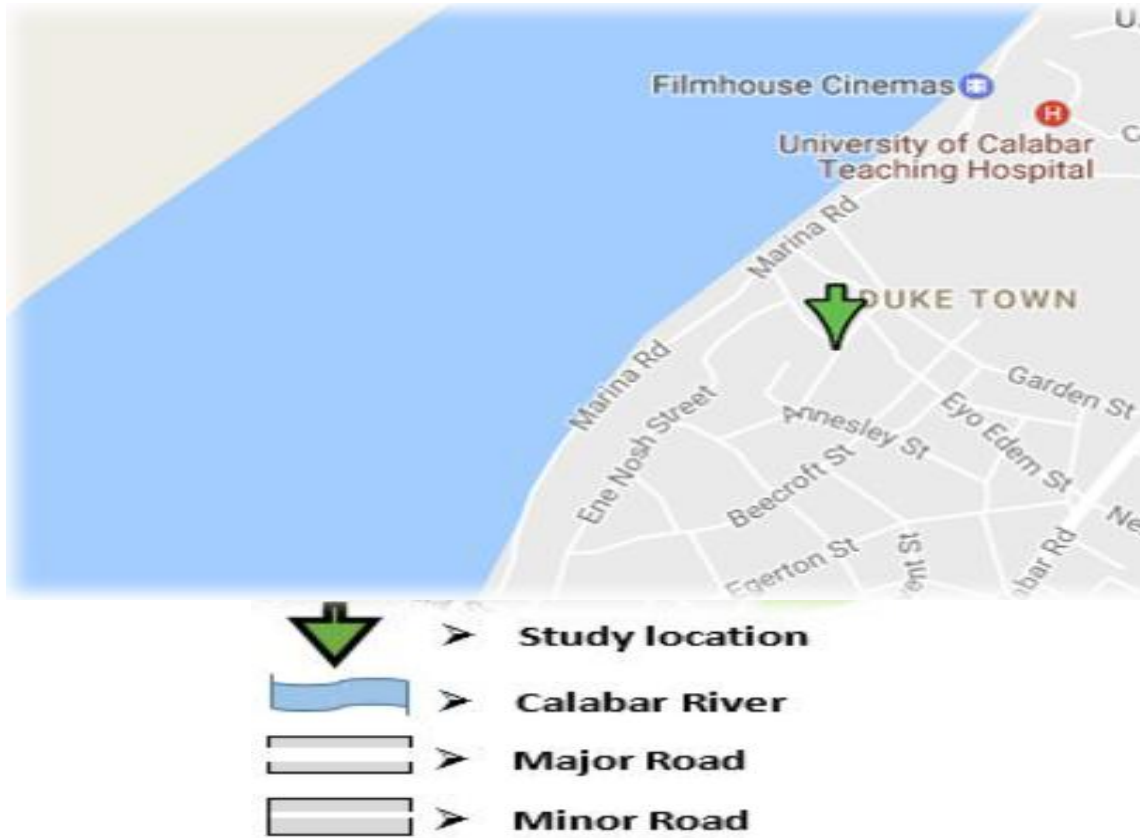

**Fig. 1:** Map showing study location

## MATERIALS AND METHODS
### MATERIALS

- A 130 watt polycrystalline solar panel with dimension of 1480*670*35mm and capacities of 7.18A and 18.10V at maximum current and voltage respectively. Having 7.91A and 21.72V for short circuit current and open circuit voltage respectively.
- Charge controller
- 12 volts lead acid battery (75AH)
- A digital thermometer capable of displaying temperature both in Celsius (C) and Fahrenheit (F) (KT-908)
- A digital hygrometer (KT-908)
- A digital alarm clock (KT-908)
- A digital multimeter (M890C$^+$) capable of measuring voltage and current, and also fitted with a K type thermocouple for measuring temperature in Celsius

### METHOD

The solar panel was placed horizontally flat facing the sun on a platform one metre high above the ground. Connecting cables were connected to the output terminals of the solar panel. From the output terminals of the solar panel the cables were connected to the charge controller. The output of the charge controller was then connected to the battery for charging the battery which powered the load through an inverter.

Measurements were taken at an interval of 30 minutes from 6.00am to 6.00pm for a period of 90 days (30 days in April and 30 days in May). During measurements, the voltage and the current from the panel were measured using the digital multimeter. The ambient temperature was read directly from the digital thermometer while the solar panel temperature was measured using the temperature sensing probe fixed on the solar panel. The time of day was recorded and the relative humidity measured and read directly from the digital hygrometer.

From the readings obtained, the power from the solar panel was determined using equation (1). The maximum power and the normalized power output efficiency was calculated using equation (2) and (3). The open circuit voltage and short circuit current depend on parameters like solar irradiance and temperature as shown in equations (4) and (5).

**Measured Power:**
$$P_{mea} = V_{mea} \times I_{mea} \tag{1}$$

**Maximum power:**
$$P_{max} = V_{max} \times I_{max} \tag{2}$$

**Normalized power output efficiency:**
$$\eta_p = \frac{P_{mea}}{P_{max}} \times 100 \tag{3}$$

**Open circuit voltage:**

$$V_{oc} = \frac{KT}{Q} \ln \frac{Isc}{Io} \qquad (4)$$

**Short circuit current:**

$$Isc = bH \qquad (5)$$

Where $P_{mea}$, $V_{mea}$ and $I_{mea}$ are the measured power, voltage and current respectively. $P_{max}$, $V_{max}$ and $I_{max}$ are the maximum power, voltage and current respectively that the module can give out. $I_o$ is the saturation current, Q is the electronic charge, K is the Boltzmann constant, T is the absolute temperature of the photovoltaic module, H is the incident light intensity and b is a constant depending on the properties of the semiconductor junction. Fig. 2 shows how the experimental setup of the photovoltaic system was done.

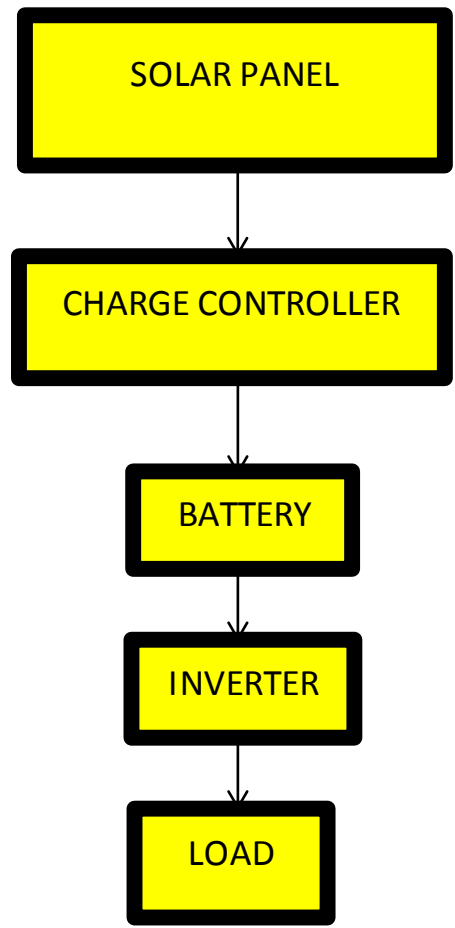

**Fig. 2: Block diagram of the experimental setup**

# RESULTS

## TABLE 1: Averages of data obtained

| Time of day (min) | Relative Humidity (%) | Ambient Temperature ($^0$C) | Panel Temperature ($^0$C) | Voltage (V) | Current (A) | Power (W) | Efficiency (%) |
|---|---|---|---|---|---|---|---|
| 6:00 | 83.36 | 27.77 | 25.65 | 3.08 | 0.00 | 0.00 | 0% |
| 6:30 | 86.27 | 27.29 | 25.20 | 14.37 | 0.09 | 1.28 | 1% |
| 7:00 | 85.53 | 27.55 | 26.20 | 16.57 | 0.31 | 5.06 | 4% |
| 7:30 | 85.00 | 27.60 | 27.79 | 16.96 | 0.58 | 9.84 | 8% |
| 8:00 | 81.29 | 28.55 | 30.54 | 17.79 | 1.39 | 24.66 | 19% |
| 8:30 | 79.75 | 29.26 | 33.39 | 18.00 | 2.02 | 36.34 | 28% |
| 9:00 | 75.13 | 30.17 | 36.29 | 17.93 | 2.39 | 42.87 | 33% |
| 9:30 | 74.06 | 30.51 | 38.18 | 17.93 | 2.71 | 48.67 | 37% |
| 10:00 | 71.53 | 31.26 | 40.18 | 17.83 | 3.58 | 63.81 | 49% |
| 10:30 | 71.41 | 31.67 | 40.65 | 18.04 | 3.21 | 57.87 | 45% |
| 11:00 | 68.29 | 32.38 | 43.50 | 17.94 | 3.77 | 67.72 | 52% |
| 11:30 | 65.71 | 32.94 | 43.06 | 18.16 | 4.27 | 77.54 | 60% |
| 12:00 | 62.47 | 33.84 | 47.12 | 17.81 | 5.07 | 90.33 | 69% |
| 12:30 | 61.65 | 34.25 | 46.18 | 17.93 | 4.90 | 87.84 | 68% |
| 13:00 | 61.71 | 33.92 | 44.03 | 17.83 | 3.85 | 68.71 | 53% |
| 13:30 | 61.18 | 34.32 | 44.59 | 17.57 | 4.22 | 74.16 | 57% |
| 14:00 | 59.06 | 35.05 | 47.46 | 17.94 | 4.36 | 78.21 | 60% |
| 14:30 | 60.00 | 35.01 | 44.58 | 17.85 | 4.21 | 75.10 | 58% |
| 15:00 | 61.41 | 34.34 | 41.76 | 17.86 | 3.33 | 59.46 | 46% |
| 15:30 | 60.88 | 34.52 | 42.73 | 17.87 | 3.13 | 55.86 | 43% |
| 16:00 | 63.88 | 33.89 | 39.28 | 17.70 | 2.53 | 44.71 | 34% |
| 16:30 | 66.29 | 33.28 | 36.95 | 17.47 | 1.77 | 30.86 | 24% |
| 17:00 | 70.24 | 32.11 | 33.51 | 17.15 | 0.88 | 15.04 | 12% |
| 17:30 | 72.06 | 31.37 | 31.93 | 17.06 | 0.60 | 10.24 | 8% |
| 18:00 | 74.06 | 30.68 | 30.08 | 16.20 | 0.24 | 3.96 | 3% |

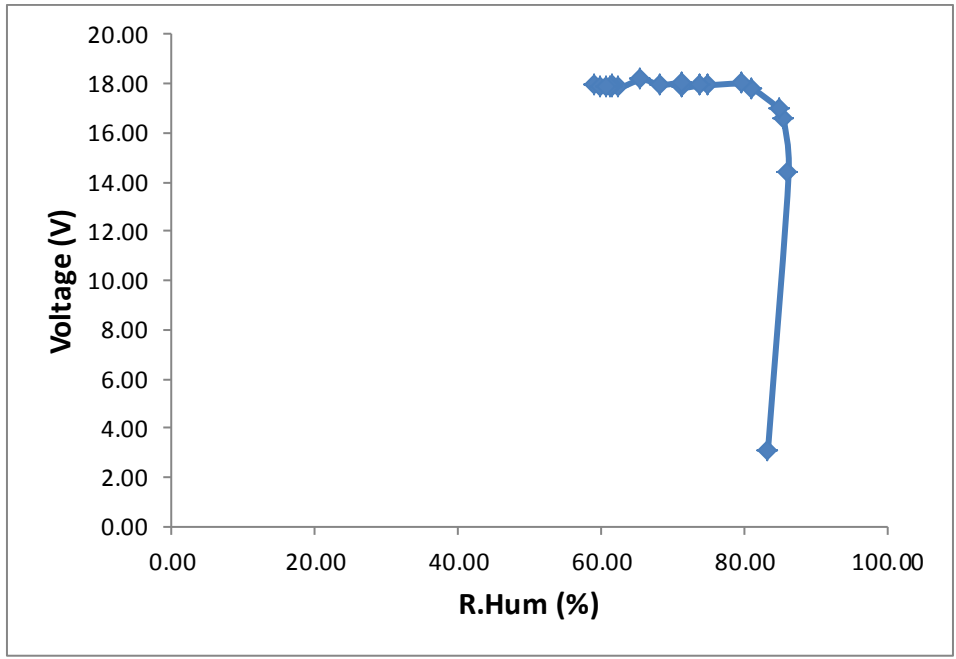

**Fig. 3: Graph of voltage against humidity**

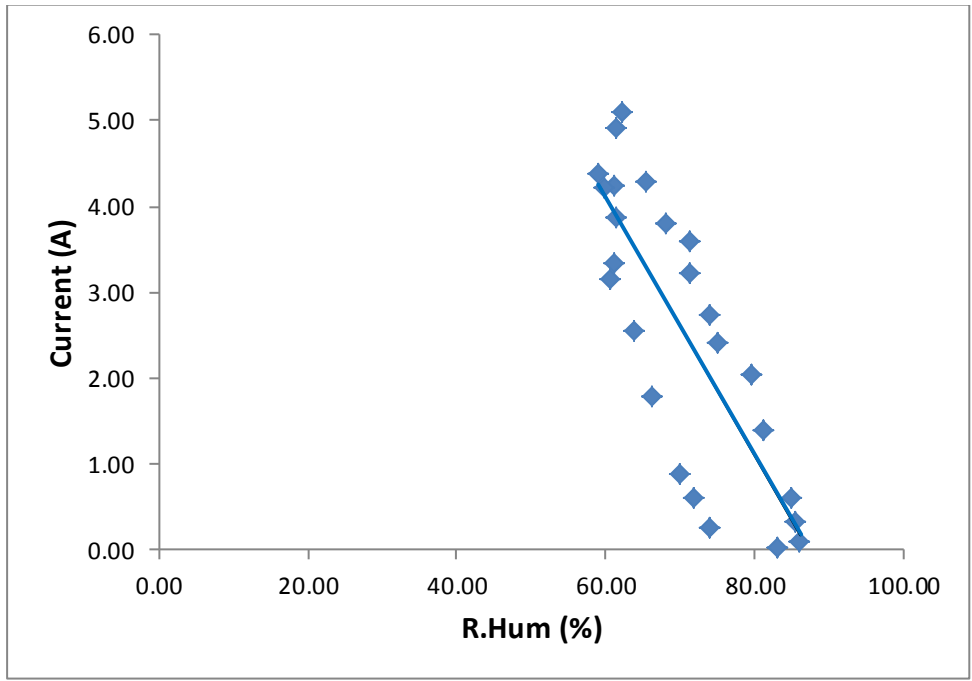

**Fig. 4: Graph of current against humidity**

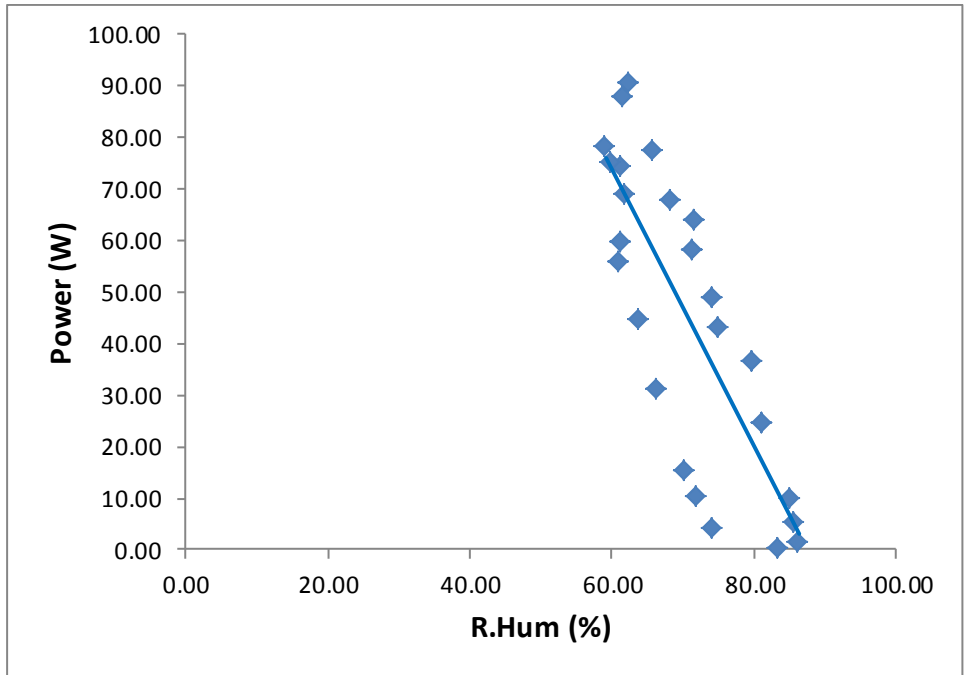

**Fig. 5: Graph of power against humidity**

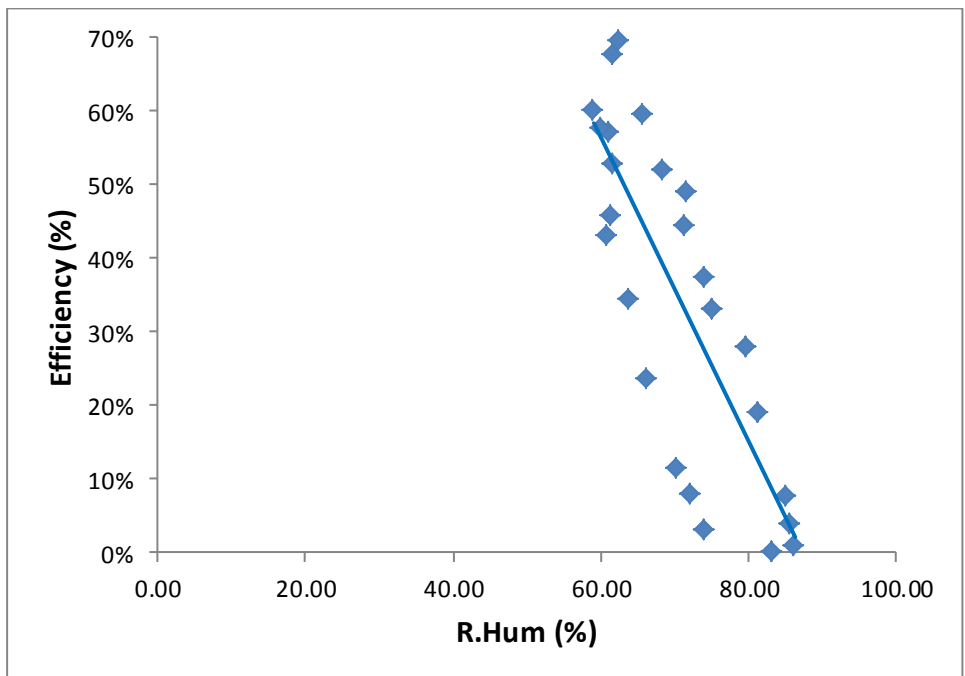

**Fig. 6: Graph of efficiency against humidity**

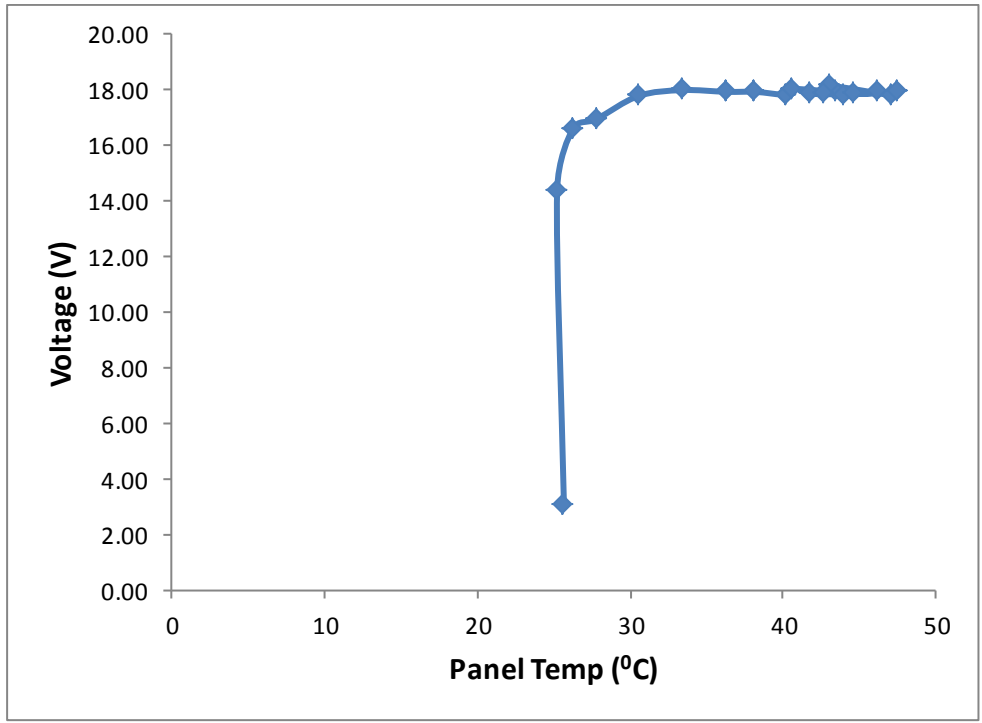

**Fig. 7: Graph of voltage against panel temperature**

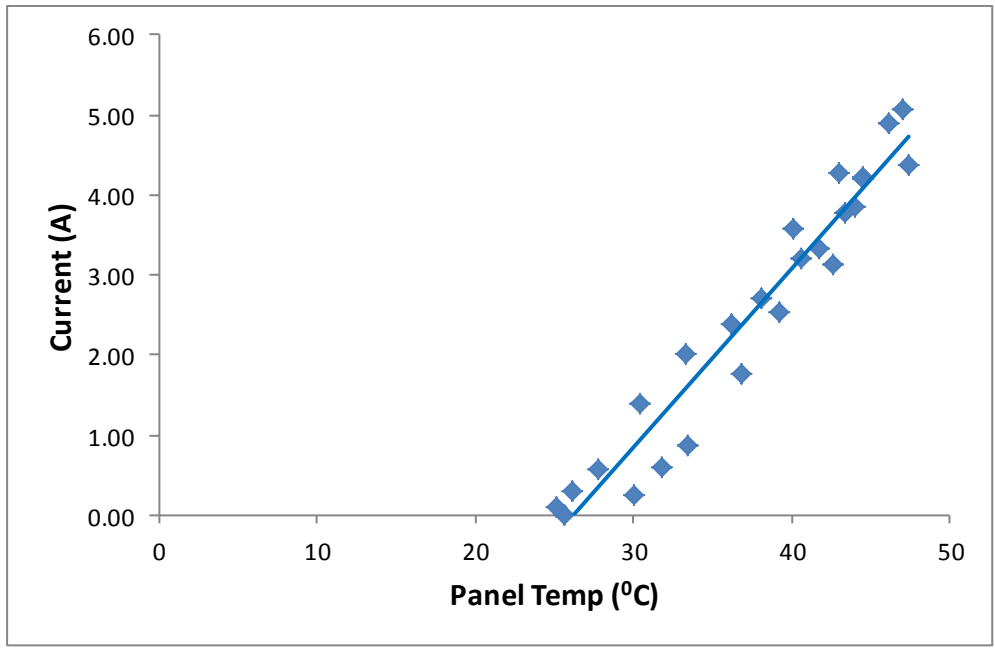

**Fig. 8: Graph of current against panel temperature**

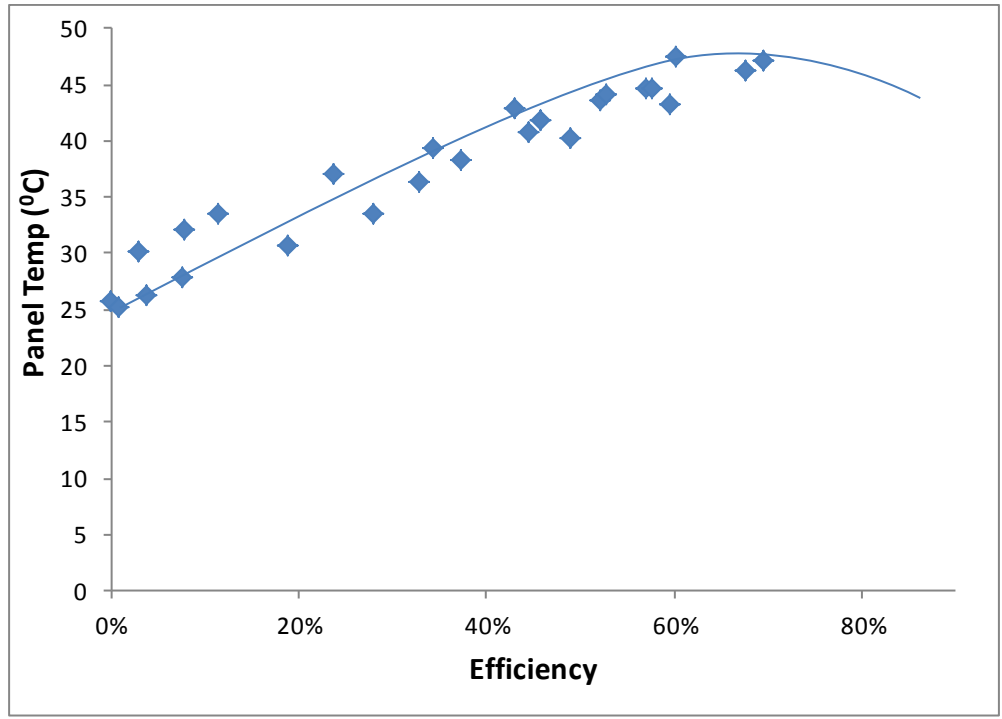

Fig. 9: Graph of panel temperature against efficiency

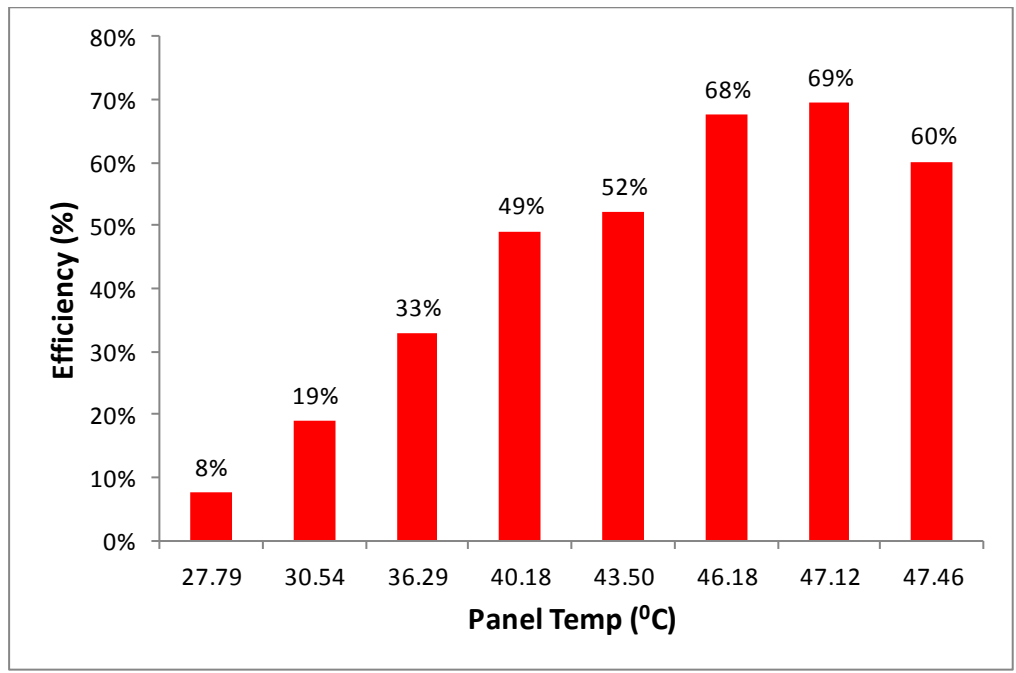

Fig. 10: Bar chart showing efficiency against panel temperature

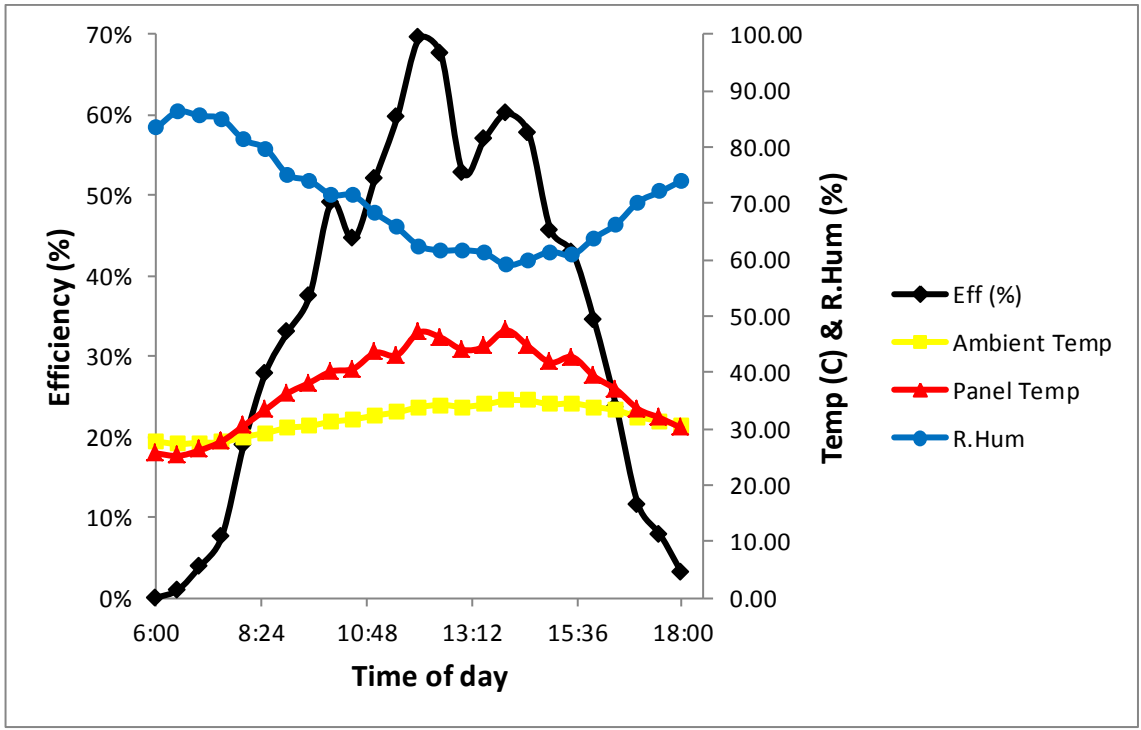

**Fig. 11: Graph of efficiency against time of day on the primary axis and graph of temperature and humidity against time of day on the secondary axis.**

## DISCUSSIONS

Table 1 show the recorded data over the period of measurement and it is from Table 1 that Figures 3 to 11 were plotted from.

Fig. 3 show that voltage increases as relative humidity decreases. The voltage remains fairly stable between 65% to 75%. Beyond 75% the voltage begins to fall sharply to almost zero.

Fig. 4, Fig. 5 and Fig. 6 show that as relative humidity decreases the current, power and efficiency of the module increases respectively, indicating that a linear relationship might exist between relative humidity and current, power, and efficiency.

Fig. 7 shows that between $25^0C$ and $33^0C$ voltage increases. With further increase in temperature, voltage remains fairly stable and beyond $43^0C$, voltage begins to drop indicating that temperature has significant effect on the voltage output.

Fig. 8 shows that current increases with temperature to about $47^0C$ where a peak in current is observed and beyond this temperature, current begins to drop, thus indicating the maximum operating temperature of the photovoltaic module.

Fig. 9 and Fig. 10 show that efficiency increases with temperature up to the maximum operating temperature of the photovoltaic module before starting to decrease. Indicating that high temperature above the maximum operating temperature of the module does not enhance high performance.

Fig. 11 shows that during the early hours of the morning, the module temperature is lower than the ambient temperature. This is because the solar cells are cased with glass and aluminum materials which have low temperature at room temperature. At noon the module temperature increases rapidly than the ambient temperature due to the high solar irradiation from the sun resulting to high amount of photon reaching the module. Towards the evening, the module temperature drops relatively faster than the ambient temperature due to the sun setting. Fig. 11 further reveals that ambient temperature has little or no direct effect on the module temperature.

Fig. 11 also reveal that peak in efficiency is observed at a relative humidity value of 65% and module temperature of about 47$^0$C at noon, indicating that these values of relative humidity and module temperature enhance the module performance.

## CONCLUSION

This research shows that the effect of relative humidity on current, power and efficiency are observed to be the same. The research also confirms that the ambient temperature has little or no direct effect on the module temperature.

The performance of the PV system is also affected by relative humidity and temperature. The portion of the absorbed solar radiation that is not converted into electricity gets converted into thermal energy and causes a decrease in module efficiency. This undesirable effect which leads to an increase in the PV cell's working temperature and consequently causing a drop of conversion efficiency can be minimized if a means can be devised to keep the module temperature from exceeding the maximum operating cell temperature.

The application of photovoltaic technology in the conversion of solar energy to electricity within the location under study in the months of April and May can be said to be favorable.

## Conflicts of Interest

The authors declare that there is no conflict of interest regarding the publication of this paper.

## References

[1] Hosseini, R., Hosseini, N. & Khorasanizadeh, H. (2011). An experimental study of combining a photovoltaic system with a heating system. *World Renewable Energy Congress 2011,* 8-13 May 2013.

[2] Chaichan, M. T., & Kazem, H. A. (2016). Experimental analysis of solar intensity on photovoltaic in hot and humid weather conditions. *International Journal of Scientific & Engineering Research, 7*(3), 91-96.

[3] Hirst, L. C., & Ekins-Daukes, N. J. (2011). Fundamental losses in solar cells. *Prog Photovolt Res Appl, 19*(3), 286-293.

[4] Tobnaghi, D. V. & Naderi, D. (2015). The effect of Solar Radiation and Temperature on Solar Cells Performance. *Extensive Journal of Applied Sciences,* 3(2), 39-43.

[5] Hussein, A. K., Miqdam, T. C. (2015). Effect of Humidity on Photovoltaic performance based on experimental study. *International Journal of Applied Engineering Research*, 10(23), 43572-43577.

[6] Panjwani, M. K. & Narejo, G. B. (2014). Effect of humidity on the efficiency of solar cell (photovoltaic). *International Journal of Engineering Research and General Science,* 2(4).

[7] Ike, C. U. (2013). The Effect of Temperature on the Performance of a Photovoltaic Solar System in Eastern Nigeria. *International Journal of Engineering And Science,* 3(12), 10-14.

[8] Ettah, E. B., Udoimuk, A. B., Obiefuna, J. N. & Opara, F. E. (2012). The effect of Relative Humidity on the Efficiency of Solar Panels in Calabar, Nigeria. *Universal Journal of Management and Social Sciences*, 2(3), 8-11.

[9] Omubo-pepple, V. B., Isreal-cookey, C. & Alaminokuma, G. I. (2009). Effects of Temperature, Solar Flux and Relative Humidity on the Efficient Conversion of Solar Energy to Electricity. *European Journal of Scientific Research,* 35(2), 173-180.

[10] Muhammad, A. B., Hafiz, M. A., Shahid, K., Muzaffar, A. & Aysha, M. S. (2014). Comparison of Performance Measurements of Photovoltaic Modules during Winter Months in Taxila, Pakistan. *International Journal of Photoenergy,* Volume 2014, Article ID 898414, 8 pages.